\documentclass[prd,aps,floatfix,notitlepage,prl,reprint,superscriptaddress,twocolumn,nofootinbib]{revtex4-1}
\usepackage{graphicx,graphics,amsmath,bm,natbib,multirow,hyperref,float,fancyhdr}
\usepackage[usenames,dvipsnames]{color}
\usepackage[caption = false]{subfig}
\hypersetup{colorlinks=true,citecolor=blue,linkcolor=blue}

\newcommand{\araa}{{\em Ann.\ Rev.\ Astron.\ Astrophys.}}

\newcommand{\aap}{{\em Astron.\ Astrophys.}}
\newcommand{\apjl}{{\em Astrophys.\ J.\ Lett.}}
\newcommand{\apjs}{{\em Astrophys.\ J.\ Suppl.}}
\newcommand{\mnras}{{\em Mon.\ Not.\ R.\ Astron.\ Soc.}}

\newcommand{\pasj}{{\em Publ.\ Astron.\ Soc.\ Japan}}

\newcommand{\be}{\begin{equation}}
\newcommand{\ee}{\end{equation}}

\begin{document}

\title{Anisotropic halo assembly bias and redshift-space distortions} 
\author{Andrej Obuljen}
\affiliation{Waterloo Centre for Astrophysics, Department of Physics and Astronomy, University of Waterloo, 200 University Ave W, Waterloo, ON N2L 3G1, Canada}
\author{Neal Dalal}
\affiliation{Perimeter Institute for Theoretical Physics, 31 Caroline St. North, Waterloo, ON N2L 2Y5, Canada}
\author{Will J. Percival}
\affiliation{Waterloo Centre for Astrophysics, Department of Physics and Astronomy, University of Waterloo, 200 University Ave W, Waterloo, ON N2L 3G1, Canada}
\affiliation{Perimeter Institute for Theoretical Physics, 31 Caroline St. North, Waterloo, ON N2L 2Y5, Canada}

\begin{abstract}

We study the effect of large-scale tidal fields on internal halo properties using a set of N-body simulations. We measure significant cross-correlations between large-scale tidal fields and several non-scalar halo properties: shapes, velocity dispersion, and angular momentum.  Selection effects that couple to these non-scalar halo properties can produce anisotropic clustering even in real-space. We investigate the size of this effect and show that it can produce a non-zero quadrupole similar in size to the one generated by linear redshift-space distortions (RSD). Finally, we investigate the clustering properties of halos identified in redshift-space and find enormous deviations from the standard linear RSD model, again caused by anisotropic assembly bias. These effects could contaminate the values of cosmological parameters inferred from the observed redshift-space clustering of galaxies, groups, or 21cm emission from atomic hydrogen, if their selection depends on properties affected by halo assembly bias. We briefly discuss ways in which this effect can be measured in existing and future large-scale structure surveys.

\end{abstract}
\maketitle

\section{Introduction} \label{sec:Introduction}

In the standard $\Lambda$CDM model, small seed fluctuations in the early Universe grow to form a complicated distribution of matter at present day. The key process at work is anisotropic gravitational collapse, which produces nonlinear structures like halos, filaments and sheets in a reticulated pattern called the cosmic web \cite{Bond1996}. 
The connection between matter and observed tracers like galaxies is complex and intricate, but on large scales the clustering strength of haloes is approximately linearly biased compared to the clustering of matter \cite{Kaiser_bias} (for a review see \cite{Wechsler_review,Bias_review}).

It has long been understood that the linear bias between halos and mass must depend on halo mass and redshift, in the sense that more massive halos are more strongly clustered than less massive, while for a fixed mass, halos are more clustered at earlier times \cite{Kaiser_bias}.
The \citet{Press1974} ansatz  predicts that bias depends {\em only} on mass and redshift \citep{Bond1991,Mo1996}.  
However, this model breaks down in detail, and the clustering properties of halos at fixed mass and redshift have been shown to additionally depend on other \textit{internal} halo properties including the formation time and history, spin, concentration, angular momentum, shape, etc. \cite{Faltenbacher,Hahn2007,Lazeyras:2016xfh,Mao2018}. These extra dependencies are commonly referred to as halo assembly bias, although ``secondary bias'' is a more apt term. To understand galaxy assembly bias we have to further include additional dependencies between galaxy numbers and halo properties, and between galaxy and halo properties \cite{Wechsler_review}. In this paper, we focus exclusively on halo assembly bias. This is the key effect for the clustering of groups of galaxies found using observed galaxy positions, while for galaxies and 21cm observations it is modulated by the galaxy--halo connection \cite{Wechsler_review}.

Halo assembly bias has been detected in numerical simulations \cite{Gao2005, Gao2007} and understood in theory using peaks \cite{Dalal2008}, and the tidal field \cite{Hahn2009}. 
Observationally, no convincing detection of halo assembly bias exists for cluster-sized halos \cite{Sunayama2019} or group-sized halos \cite{Lin2016}. \cite{Paranjape2018} considered the clustering of redshift-space group catalogues in the Sloan Digital Sky Survey, as a function of the local tidal field. They found that all qualitative trends in the monopole matched those in simulations with no assembly bias, suggesting that mass-dependent effects dominate the trends seen. Differences in the clustering of Luminous Red Galaxies as observed by the Baryon Oscillation Spectroscopic Survey (BOSS) have also been found after splitting the sample by star formation history \cite{Montero-Dorta2017}, and by orientation \cite{Hirata_obs}.

The picture of halo assembly bias is complicated because halo properties are themselves correlated. For example, older haloes tend to be more concentrated, and formed in stronger tidal fields. Thus, it is not immediately obvious that a correlation between any one halo property and the large-scale clustering strength means that this property is the root cause of assembly bias \cite{Mao2018}.
Nevertheless, several works have shown that tidal fields are important in producing the assembly bias observed in low-mass halos \citep{Mansfield2019,Ramakrishnan}.  

An additional complication for real data is that we observe the positions of objects in redshift-space. On large scales, the standard model for the effects of redshift-space-distortions (RSD) is the linear model of \cite{Kaiser}. This model assumes that the tracer field is related to the underlying real-space matter field via some (possibly nonlinear) transformation, and then tracers are put into redshift-space in a manner that conserves number.  This is a good description for halo cores and galaxies on large scales.  However, there are cases where this \citeauthor{Kaiser} picture is not valid, e.g.\ Ly-$\alpha$ forest \cite{Uros} and voids \cite{Chuang}, where one of the tenets of the Kaiser derivation breaks down. 

The amplitude of assembly bias is intimately connected with the selection of objects for which clustering is measured. For observational data, redshift-space effects need to be included when considering the selection of objects. \citet{Hirata_th} argued that selection linked to orientation in redshift-space will lead to an anisotropic clustering signal in addition to RSD (see also \cite{Desjacques}). The amplitude of this effect depends on the object and selection method. Both \cite{Padilla2019} \&~\cite{McCarthy2018} used simulations to investigate whether assembly bias can affect standard large-scale measurements based on galaxy clustering, and found only a small signal at the limits of their simulations, consistent with assembly bias having a weak effect on galaxies selected with a standard algorithms. This does not mean that assembly bias has to be small for groups selected from the observed positions of galaxies, as we show below.

In this paper we use simulations, described in \S\ref{sec:Simulations}, to further investigate how non-scalar internal halo properties are correlated with large-scale anisotropic tidal fields. Using these simulations we find that halo shapes, velocity dispersions, and angular momenta are highly correlated with tidal fields on large scales (\S\ref{sec:Tidal}). These forms of anisotropic assembly bias imply that any orientation-sensitive selection procedure will lead to anisotropic large-scale clustering, even in real-space. This holds true not only for selection of individual objects like galaxies, but also for selection of galaxy groups and clusters in redshift-space. In \S\ref{sec:consequences} we illustrate how various selection effects can introduce anisotropy in the real-space power spectrum, and thereby contaminate redshift-space clustering measurements of the growth of cosmic structure made with groups. We discuss the results and possible consequences in \S\ref{sec:Conclusions}.

\section{Numerical simulations}\label{sec:Simulations}

In order to investigate in detail the internal properties of halos and their correlations with the large-scale tidal field, we use 1000 N-body simulations, a subset of Quijote\footnote{\href{https://github.com/franciscovillaescusa/Quijote-simulations}{https://github.com/franciscovillaescusa/Quijote-simulations}} suite \citep{Quijote}. The subset we use are the output snapshots at $z=0$ from the dark matter only simulations run using TreePM+SPH code Gadget-III \cite{Gadget} in a periodic box size of $1\,h^{-1}{\rm Gpc}$ with $512^3$ particles. The mass of a single particle is $M_{\rm p}=6.57\times10^{11}\,[h^{-1} M_\odot]$. All simulations were run using the following values of cosmological parameters: $\Omega_{\rm m} = 0.3175$, $\Omega_{\rm b} = 0.049$, $\Omega_\Lambda = 0.6825$, $n_{\rm s} = 0.9624$ and $h = 0.6711$, which are in good agreement with the constraints from Planck \cite{planck15}.
We use the Friends-of-Friends (FoF) algorithm \cite{FoF} to identify halos both in real- and redshift-space using the linking length $b_{ll}=0.2$ and different minimum number of particles per halo $n_{\rm min}$.

\begin{figure*}[!ht]
\subfloat{\includegraphics[width=0.48\textwidth]{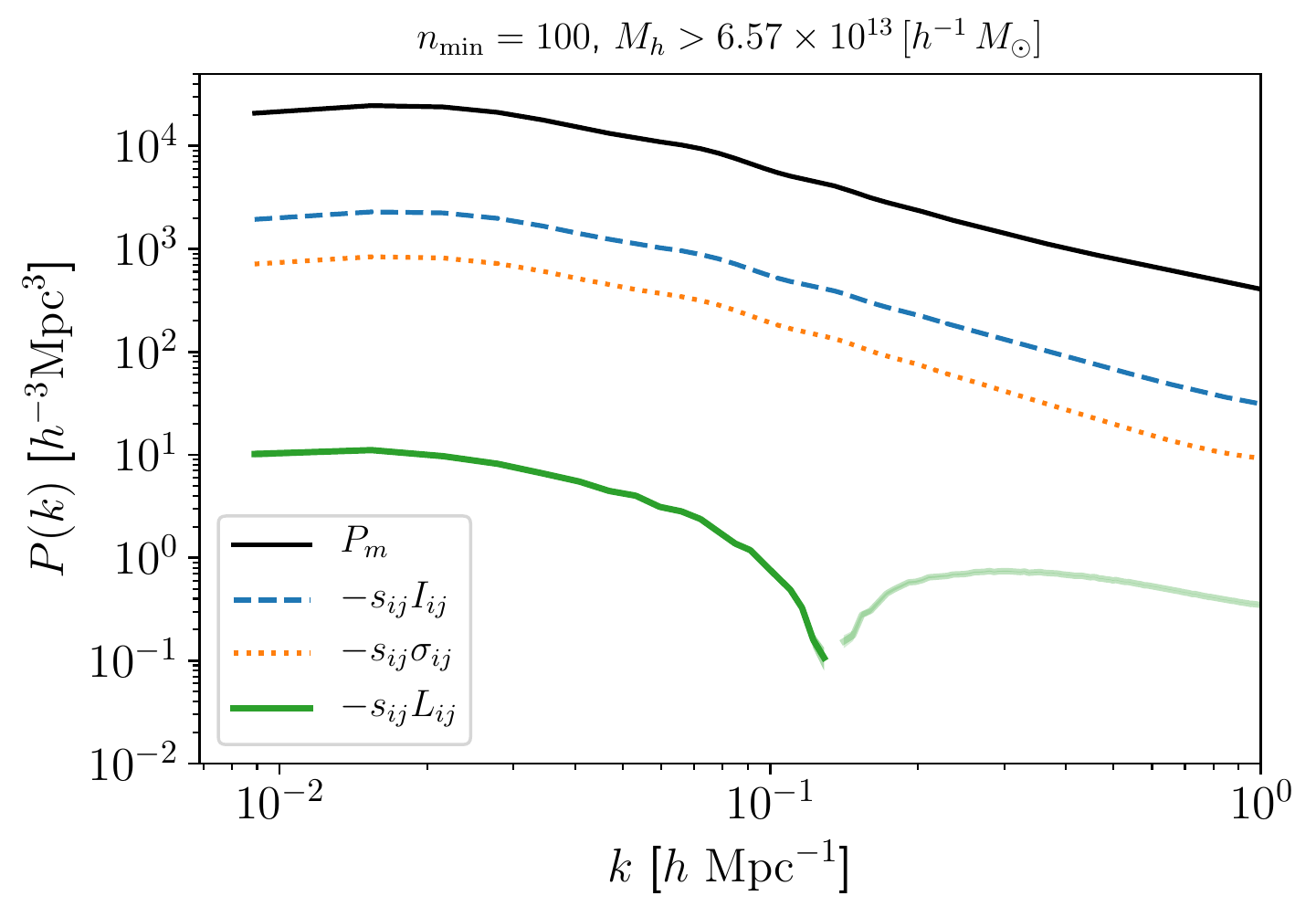}}
\subfloat{\includegraphics[width=0.48\textwidth]{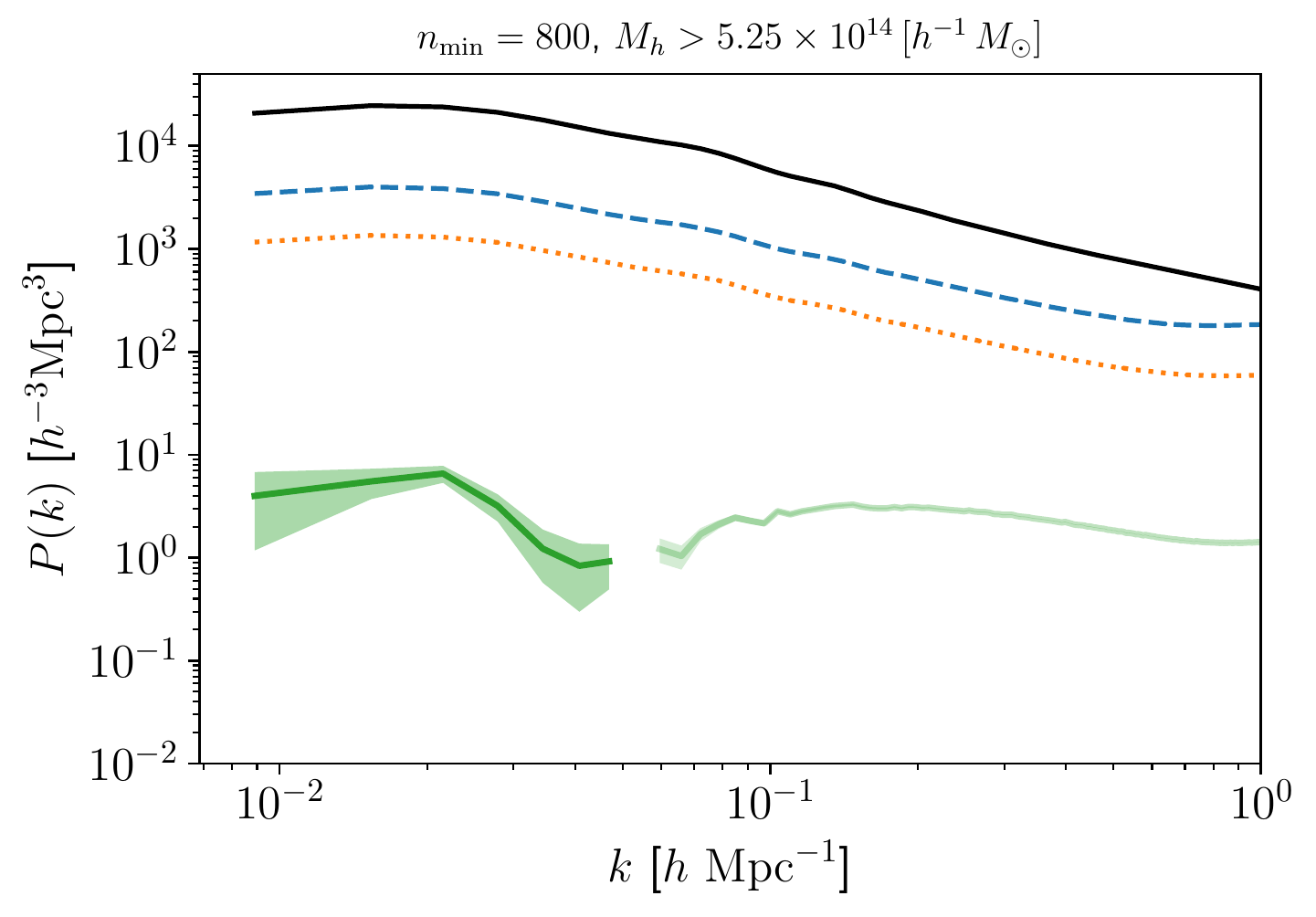}}
\caption{Real-space cross-correlations of the tidal field $s_{ij}$ with various internal halo properties: shapes ($I_{ij}$, blue dashed lines), velocity dispersion ($\sigma_{ij}$, orange dotted lines) and angular momentum ($L_iL_j$, green solid lines), for two different halo mass bins. For comparison, we also show the matter power spectrum (black solid lines). Shaded areas represent $1\sigma$ uncertainty, computed as the error on the mean from 1000 Quijote simulations. While shapes and velocity dispersion are negatively correlated with the tidal field on all scales, angular momentum shows a nontrivial scale dependence and changes sign in a mass-dependent manner.
}
\label{fig:Correlations}
\end{figure*}

\section{Tidal field correlations}
\label{sec:Tidal}

Halo bias can be understood as a consequence of correlations between properties of halos and properties of the large-scale environments of those halos.  Peaks bias is a well-known example of this: the presence of a large-scale background overdensity raises the masses of the halos arising from small-scale peaks, leading to massive halos being clustered in high-density environments \citep{Kaiser_bias}.  We normally consider only the overdensity $\delta$ associated with long-wavelength modes when studying large-scale bias, but there are other properties of long-wavelength modes that can affect halo properties and therefore lead to assembly bias. An example of this is the tidal field, which we will define as
\be \label{tide}
s_{ij}(k)=\left(\frac{k_i k_j}{k^2} - \frac{\delta_{ij}}{3}\right) \delta_m(k),
\ee
where $\delta_m(k)$ is the matter overdensity field, and $\delta_{ij}$ is the Kronecker $\delta$.  Since this is a traceless tensor quantity, by symmetry $s_{ij}$ cannot correlate with any scalar property of a halo like mass, concentration, etc.  However, it can correlate with non-scalar halo properties.  A well-known example of this is the GI correlation between (galaxy) shapes and the large-scale tidal field, which acts as a contaminant for weak gravitational lensing \cite{Hirata2004}.  For halos, the analogue would be a correlation between $s_{ij}$ and the shape of a halo.  We will define halo shapes using the dimensionless quadrupole tensor,  
\be
I_{ij} = \frac{1}{N_p}\sum_{k=1}^{N_p} \Delta x_{k,i} \Delta x_{k,j},
\ee
where $N_p$ is the number of particles in a given halo, and $\Delta x_{k,j}$ is the $j$-component of the unit vector $\Delta\hat{\bm{x}}_k$ pointing from the halo center to the position of the $k^{\rm th}$ particle.  To quantify the correlation between $s_{ij}$ and $I_{ij}$, we first define a field $I_{ij}(\bm{x})$ from the shape tensors of individual halos as 
\be
I_{ij}(\bm{x}) = \frac{\sum_h W_h(\bm{x}) I_{h,ij}}{\sum_h W_h(\bm{x})},
\ee
where the weight function is normalized to have unit integral, 
$\int d^3 x W_h(\bm{x})=1$.  For concreteness, we use triangular-shaped cloud deposition \cite{HockneyEastwood} for $W_h$. We then Fourier transform and define the cross-power spectrum
\be
P_{\rm cross}(k) = {\rm Re}\left\langle \sum_{ij} s_{ij}^*(k) I_{ij}(k)\right\rangle.
\ee

We can similarly measure correlations between other halo properties and tidal fields.  Besides the dimensionless shape tensor, we also define a dimensionless velocity dispersion tensor 
\be
\sigma_{ij} = \frac{1}{N_p}\sum_{k=1}^{N_p} \Delta v_{k,i} \Delta v_{k,j},
\ee
where $\Delta v_{k,j}$ is the $j$-component of the unit vector $\Delta\hat{\bm{v}}_k$ pointing along the relative velocity between particle $k$ and its host halo.  Finally we also define the dimensionless angular momentum vector as
\be
L_i = \frac{1}{N_p}  \sum_k \epsilon_{ijl} \Delta x_j \Delta v_l,
\ee
where $\epsilon_{ijl}$ is the antisymmetric Levi-Civita tensor.  
We use unit vectors in defining $I_{ij}$, $\sigma_{ij}$, and $L_i$, to facilitate the comparison between the strengths of their  correlations with the large-scale tidal field.  

We next compute cross-power spectra between the tidal tensor $s_{ij}$ and halo shapes $I_{ij}$, velocity dispersion $\sigma_{ij}$, and angular momenta $L_i L_j$. 
Fig. \ref{fig:Correlations} shows the resulting cross-power spectra for various halo mass bins.  We find that all of these quantities are significantly correlated with tides across all scales.  The shape tensor and dispersion tensor are anti-correlated with $s_{ij}$ using our sign convention for the tidal field, and appear to linearly trace $s_{ij}$ on large scales, with bias coefficients typically a factor $\sim 2-3\times$ larger for shapes than for dispersion.  The cross-spectrum between $s_{ij}$ and $L_i L_j$ has a more interesting scale dependence.  On large scales, $L_i L_j$ anti-correlates with $s_{ij}$, but on small scales these quantities positively correlate.  If we try to write $L_i L_j$ as a biased tracer of $s_{ij}$ then the bias exhibits scale dependence over surprisingly large scales, much larger than the Lagrangian sizes of the corresponding halos.  We should note that this scale-dependence only hold for relatively large halo masses, $M \gtrsim 10^{13} M_\odot$.  Our simulations do not resolve halos of lower mass, however we have studied halos with $M_{\rm vir} < 10^{12} M_\odot$ from the  
$\nu^2$GC-L simulation \cite{Ishiyama2015}, for which Rockstar halo catalogs \cite{Behroozi2013} are publicly available.\footnote{See \url{ http://hpc.imit.chiba-u.jp/~nngc/data.html}}  
The Rockstar catalog lists the angular momentum vector $L_i$ for each identified (sub)halo, and so we compute the cross-correlation of the $L_i L_j$ field with the large-scale tidal field as determined using these (sub)halos as a proxy for mass.  For these low-mass halos, the cross-spectrum does not exhibit the strong scale dependence at $k \lesssim 0.1 h\,{\rm Mpc}^{-1}$ that we find for massive groups and clusters.

The correlation between halo shapes and large-scale tides is not surprising, since it is the halo version of the well-known tidal alignment effect that gives rise to the GI correlation from weak lensing \cite{Hirata2004,Hirata2007}.  The sign of this correlation is easy to understand from the theory of ellipsoidal collapse \citep{BondMyers1996}.  Under our sign convention for $s_{ij}$ in Eqn.\ \eqref{tide}, a positive eigenvalue corresponds to a direction where tides act to compress the flow of dark matter.  The directions with the largest eigenvalues tend to collapse earliest, while directions with the smallest (most negative) eigenvalues collapse later.  In collapsed halos, the axes that collapse earliest correspond to the minor axes with the smallest eigenvalues of $I_{ij}$, while the late-collapsing axes become major axes with the largest eigenvalues of $I_{ij}$.  Therefore, the eigenvalues of $s_{ij}$ anti-correlate with the eigenvalues of $I_{ij}$, leading to a negative cross-power spectrum.  The same argument also explains the sign of the cross-correlation between $s_{ij}$ and $\sigma_{ij}$: in collapsed halos, motions of DM particles are largest along the long axes and smallest along the short axes (e.g., \citep{Lithwick2011}).  The complicated shape of the $L_i L_j$ cross-correlation would appear to preclude a simple explanation, but the magnitude of the correlation is smaller than the shape and dispersion correlations by about a factor of $\lambda^2$, where $\lambda \sim 0.05$ is a typical value for the halo spin parameter \citep{Maccio:2006wpz}.

\section{Consequences}
\label{sec:consequences}

The existence of significant correlations between non-scalar properties of halos and large-scale tidal fields implies that halos exhibit anisotropic assembly bias.  This anisotropic bias can have important consequences for the observed clustering of halos and galaxies in redshift-space, as noted by \cite{Hirata_th, Hirata_obs}.  One way to see this is to consider the line-of-sight (LOS) component of the tidal tensor, which in Fourier space is $s_{zz}(k) = (\mu^2 - 1/3) \delta_m(k)$, where $\mu = k_z/k$.  The angular factor $\mu^2 - 1/3$ is precisely the same as the angular weighting $P_2(\mu)$ used to compute the quadrupole moment of the 2-point function in redshift-space.  Therefore, any nonzero correlation with the tidal field is degenerate with RSD, which can complicate the interpretation of measurements in redshift-space. Care therefore needs to be taken when using RSD to measure the growth rate of cosmic structure (through $f=d\log D/d\log a$) if using an object selection algorithm correlated with tides, which could lead to significant errors and biases in measurements \cite{Hirata_th, Hirata_obs}. In general, to date, RSD have been measured from samples selected such that such correlations are averaged over and are therefore small \cite{Padilla2019,McCarthy2018}. Below we illustrate potential biases that could arise with specific examples.

\subsection{Quadrupole in real-space}
\label{sec:Realspace}

\begin{figure}
\includegraphics[width=0.48\textwidth]{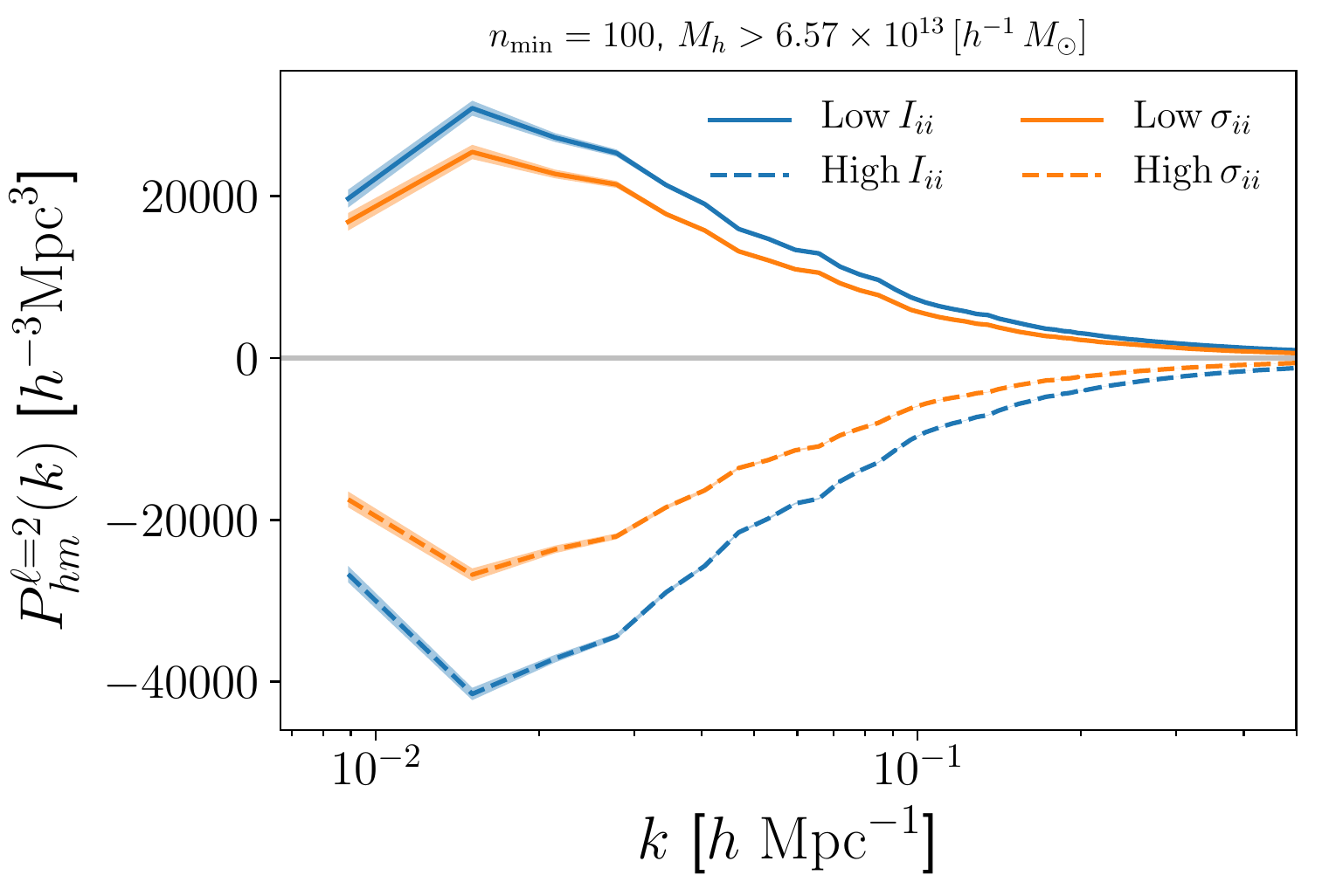}
\includegraphics[width=0.48\textwidth]{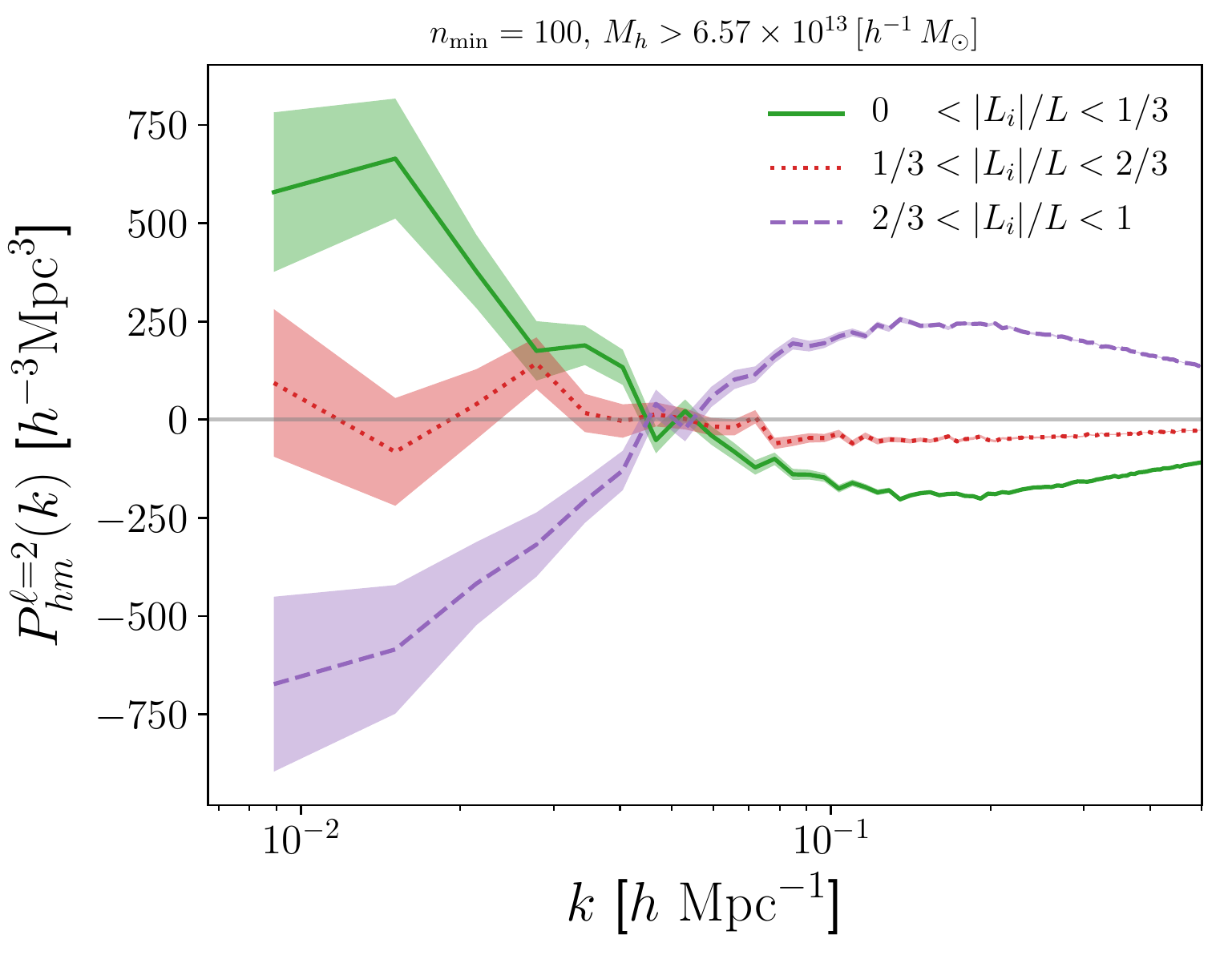}
\caption{Measured quadrupole of the halo-matter cross power spectrum in real-space for various halo selections using internal halo properties. Top panel shows the quadrupole for halos selected based on their LOS extent (blue) and velocity dispersion (orange). The bottom panel shows results for halo selection based on orientation of the net halo angular momentum.  The shaded areas in both panels are the $1\sigma$ uncertainties computed as the error on the mean over all simulations.}
\label{fig:Split}
\end{figure}

In order to reveal anisotropic halo bias, we measure the Legendre moments of the power spectrum with respect to the LOS, taken to be along one axis of the simulation
\be
P_{hm}^\ell(k) = \frac{2\ell+1}{2}\int_{-1}^{+1} P_{hm}(k,\mu) P_\ell(\mu) d\mu,
\ee
where $P_\ell(\mu)$ is the Legendre polynomial of degree $\ell$, with argument $\mu = k_\parallel / k$.
We repeat this procedure with the LOS taken along each of the three coordinate axes and compute the average multipoles for $\ell=0$ (monopole) and $\ell=2$ (quadrupole). In Fig. \ref{fig:Split} we show the resulting quadrupoles (in real-space) for the halo populations with mass $M_h>6.57\times10^{13}[h^{-1}M_\odot]$, split by internal halo properties. For the LOS extent and velocity dispersion we compare the top and bottom 10\%, while for the net halo angular momentum we split in direction. The errorbars have been computed as the error on the mean over all simulations for a single axis. Note that all of the quantities used for selection are normalized dimensionless quantities. Thus the selection of top values of, e.g.\ velocity dispersion, will not be dominated by massive halos which have higher values of $\sigma$. 

Fig. \ref{fig:Split} shows that this type of real-space selection introduces a nonzero quadrupole. Of the three halo properties we examined, the effect is strongest for the split on halo shape, as expected from magnitudes of the tidal field correlations shown in Fig. \ref{fig:Correlations}. Splitting on LOS velocity dispersion results in a similar (though weaker) effect. 
Splitting halos based on the direction of their angular momentum vectors also results in a highly significant nonzero quadrupole, though with a much smaller amplitude than what was obtained for the shape and dispersion-based splits. This is again consistent with our previous finding that the correlation of tides and $L_iL_j$ is weaker than the correlations with shapes and velocity dispersions.  Note that the scale-dependence of the $s_{ij}L_iL_j$ cross correlation appears different than the scale dependence shown in Fig.\ \ref{fig:Split}; e.g.\ the zero-crossing occurs at larger scale in Fig.\ \ref{fig:Split} than in Fig.\ \ref{fig:Correlations}.  This is because in Fig.\ \ref{fig:Correlations} we correlate tides with the angular momentum vector $\bm{L}$, while in Fig.\ \ref{fig:Split} we split based only on the direction of $\bm{L}$, i.e.\ using the unit vector $L_i/|L|$.  If we instead split halos based on the magnitude of $L_i$, then the quadrupole shows similar scale dependence as in Fig.\ \ref{fig:Correlations}, and conversely, if we compute the cross-spectrum between the tidal field $s_{ij}$ and the unit vectors $L_iL_j/L^2$ then the scale dependence of the cross-spectrum resembles Fig.\ \ref{fig:Split}.   We find similar behaviour for other halos masses as well, with the zero-crossing shifting to larger scales with higher halo masses. 

One important point to stress is that the size of the real-space quadrupole shown in Fig.\ \ref{fig:Split} is similar to the size of the quadrupole generated by the transformation from real-space to redshift-space, at least for halo shapes $I_{ij}$ and dispersion $\sigma_{ij}$.  Therefore, we can expect ${\cal O}(1)$ effects on the RSD for samples selected on these properties, as we discuss next.

\subsection{Redshift-space distortions}
\label{sec:Redshiftspace}

\begin{figure*}
\subfloat{\includegraphics[width=0.48\textwidth]{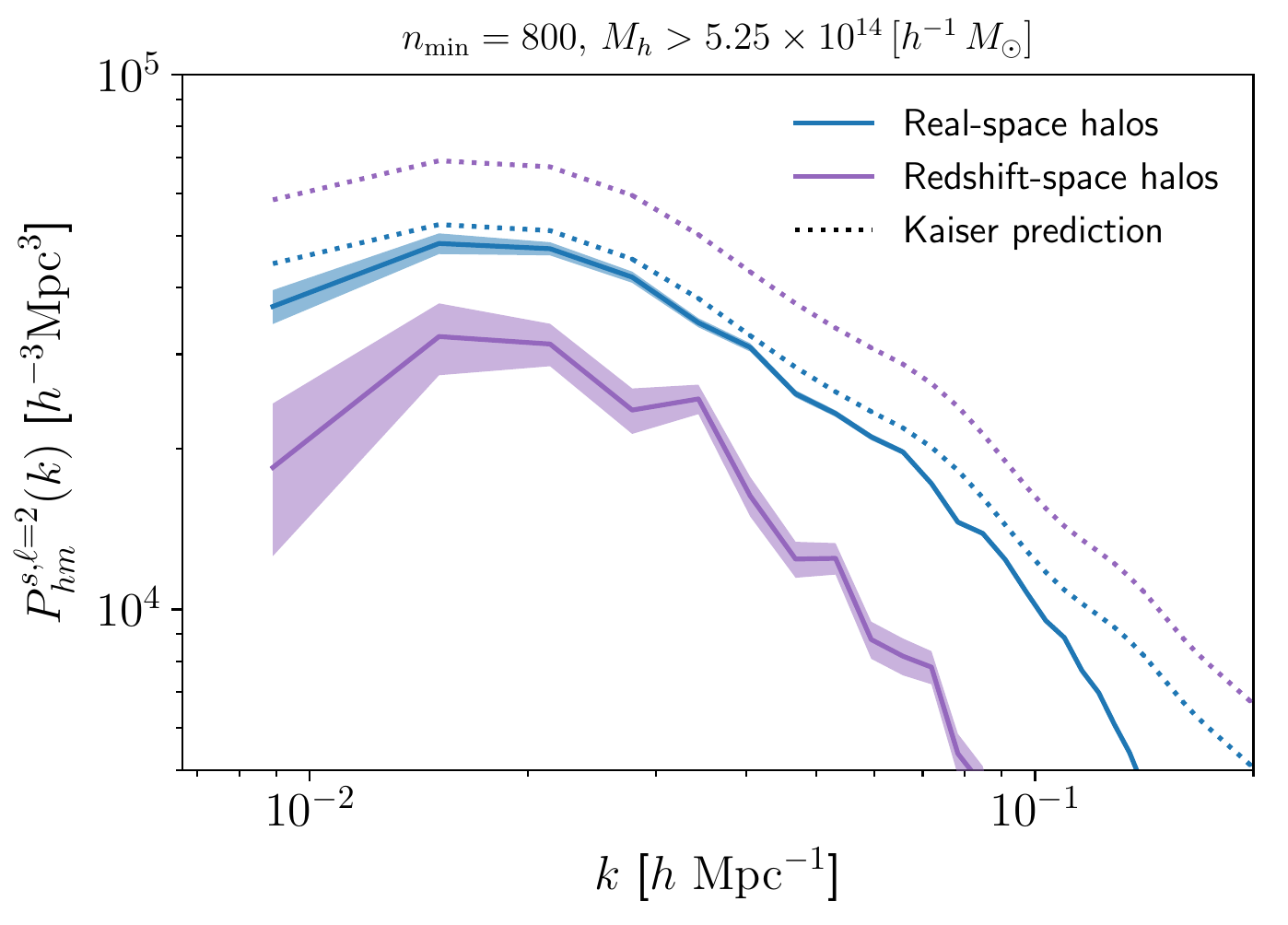}}
\subfloat{\includegraphics[width=0.48\textwidth]{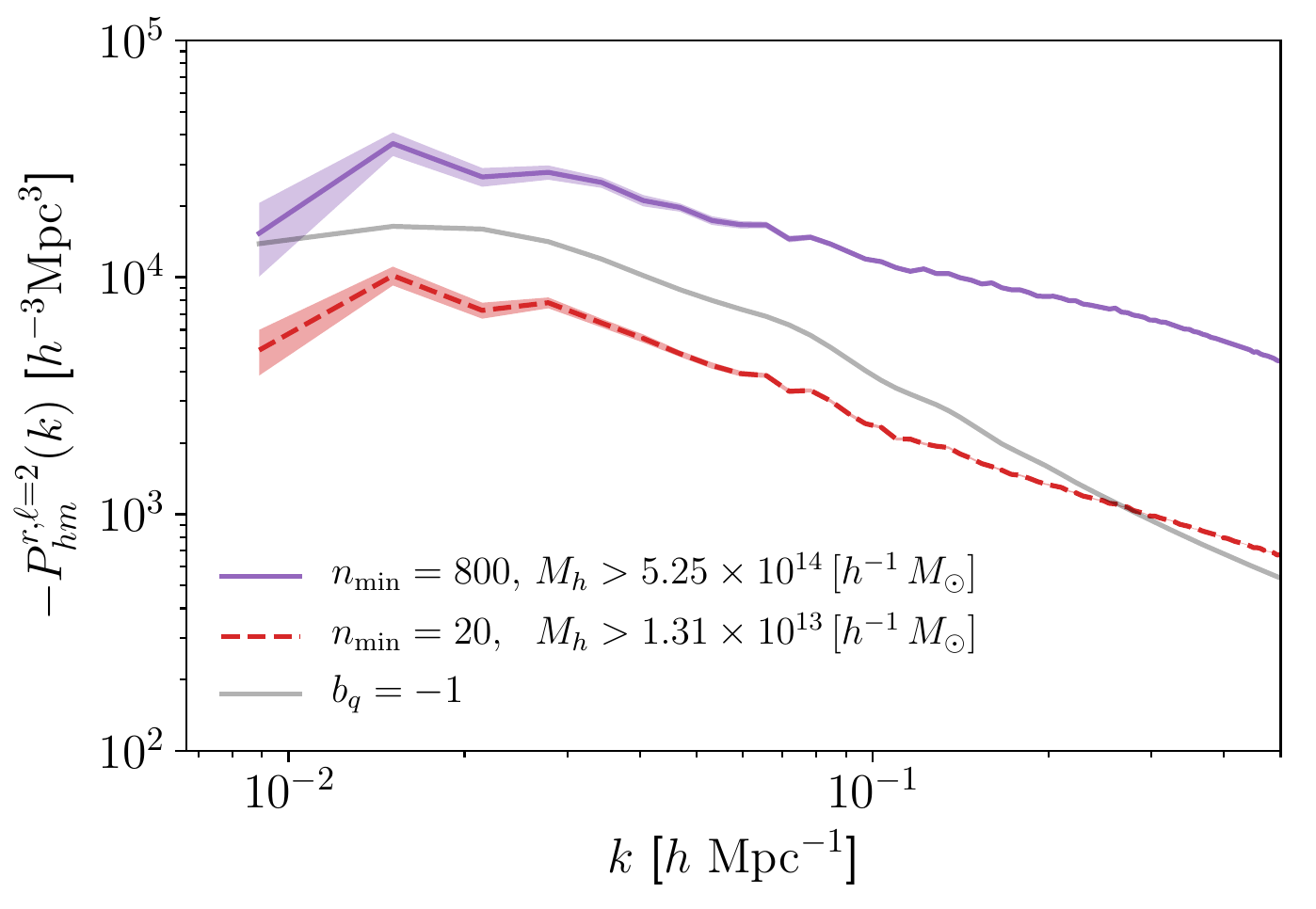}}
\caption{\textit{Left}: Quadrupole of the halo-matter cross power spectrum in redshift-space in the case halos are found in real- versus redshift-space. We also show the corresponding Kaiser quadrupole predictions (dashed lines) where we use the known value of $f$ and fit to the amplitude of the monopole on large-scales to estimate the bias. \textit{Right}: Quadrupole of the cross power spectrum in real-space of particles that end up in halos with different masses found in redshift-space with the matter field in real-space. For comparison, we show the expected real-space quadrupole in the case of vanishing Kaiser RSD ($f=0$) and $b_q=-1$ (gray solid line). Note the negative sign of the measured quadrupole. In both panels the lines and shaded regions represent the mean and the $1\sigma$ uncertainty on the mean over 1000 realizations.}
\label{fig:Quadrupoles}
\end{figure*}

On large scales in linear theory the effect of RSDs is described by the Kaiser model \cite{Kaiser}:
\be
\delta_g(\bm{k}) = (b_g + f\mu^2) \delta_m(\bm{k}),
\label{eqn:Kaiser_RSD}
\ee
where $\delta_m$ is the (real-space) matter overdensity, $\delta_g$ is the (redshift-space) overdensity of tracers, 
$b_g$ is the linear bias of the tracers, $f=d \ln D(a)/d \ln a$ is the logarithmic growth rate, and $D(a)$ is the linear growth factor. As Eqn.\ \eqref{eqn:Kaiser_RSD} makes explicit, when the matter field $\delta_m$ is statistically isotropic in real space, then redshift-space distortions are the only source of anisotropy in observed 2-point statistics, so that the observed quadrupole is proportional to $f$. When halos are selected on non-scalar quantities, there can be an extra correction to the quadrupole due to the physical correlation with tidal fields that adds to the Kaiser distortion, of the form \cite{Hirata_th,Hirata_obs}
\be
\delta_g(\bm{k}) = \left[b_g + b_q \left(\mu^2 - \frac{1}{3}\right) + f\mu^2\right] \delta_m(\bm{k}).
\label{eqn:notKaiser_RSD}
\ee
If RSD for tracers with nonzero $b_q$ are modeled using Eqn.\ \eqref{eqn:Kaiser_RSD}, this extra correction to the quadrupole from anisotropic assembly bias will lead to incorrect determination of $f$. This correction arises for selection effects that couple to the tidal field, and does not necessarily allow for all possible selection based effects \cite{Chuang}.

As discussed above, because galaxies are necessarily observed in redshift-space, any group-finding algorithm will generically be sensitive to non-scalar properties of groups that correlate with tidal fields, introducing additional corrections to the observed quadrupole beyond the Kaiser RSD. As well as these effects, which depend on a physical correlation due to halo assembly bias, we will also see apparent anisotropic clustering due to the redshift-space selection \cite{Chuang}. We illustrate the combined effect with the example of FoF groups found in redshift-space.  We use the same isotropic linking length ($b_{ll}=0.2$) used to find real-space halos, and then measure the quadrupole of the halo-matter cross-spectrum $P_{hm}$.  In redshift-space, the observed quadrupole significantly deviates from the standard Kaiser prediction on large scales, as shown in Fig.\ \ref{fig:Quadrupoles}.  The Kaiser prediction is estimated using the measured monopole of the cross-spectrum, which (knowing $f$) determines the bias $b$ for this sample.  This disagreement with the Kaiser prediction occurs for objects selected in redshift-space, but not objects selected isotropically in real-space (see blue curves in Fig.\ \ref{fig:Quadrupoles}, for example).
Another way to illustrate the discrepancy with the Kaiser model is to measure the real-space quadrupole of groups selected in redshift-space.  In real-space, the RSD vanish, meaning that any nonzero quadrupole is associated with selection effects or correlations with the tidal field. As the right-hand panel of Fig.\ \ref{fig:Quadrupoles} shows, the real-space quadrupole of the redshift-space groups is indeed significantly nonzero.

\begin{figure*}
\includegraphics[width=0.96\textwidth]{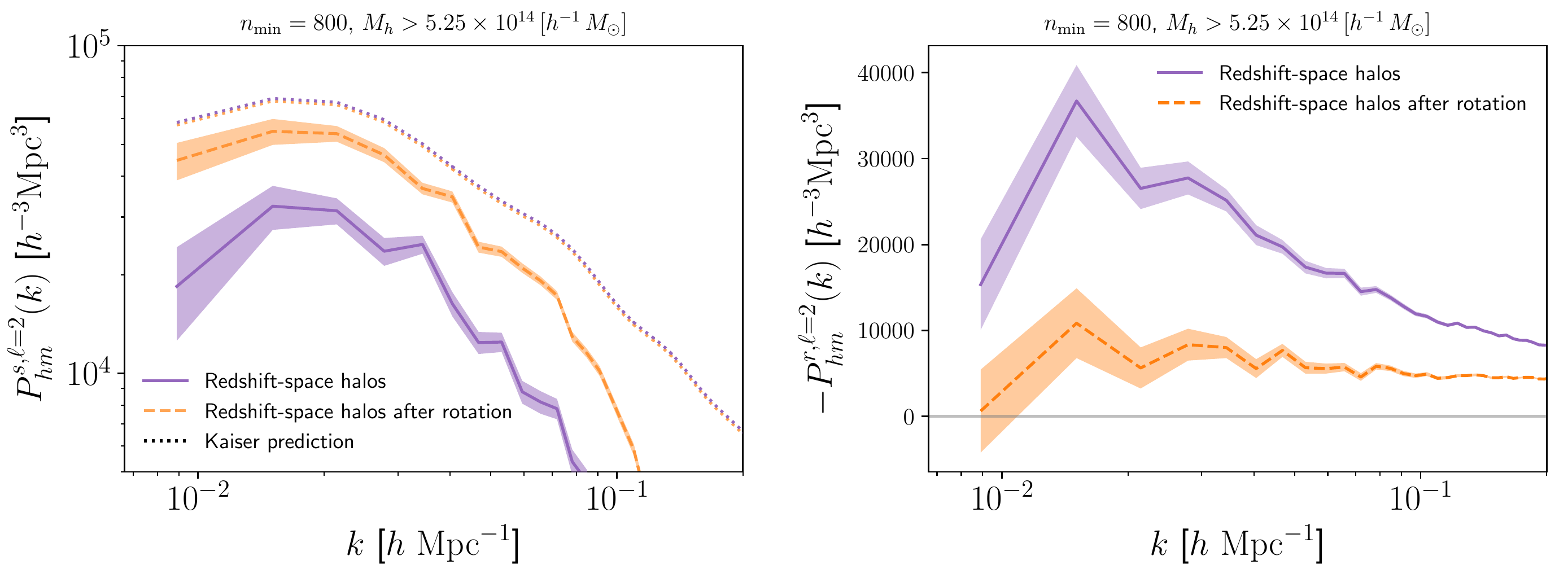}
\caption{\textit{Left}: Quadrupole of the halo-matter cross power spectrum in redshift-space in the case halos are found in redshift-space with (orange) and without (purple) rotation in real-space. Similar to Fig. \ref{fig:Quadrupoles}, we also show the corresponding Kaiser quadrupole predictions (dashed lines). \textit{Right:} Average real-space quadrupole of the cross power spectrum between matter and members of groups selected in redshift-space. For comparison, the orange curve shows the quadrupole that arises when we first randomly rotate all real-space halos before finding redshift-space groups; the difference on large scales shows that the low-$k$ quadrupole in the purple curve does indeed arise from anisotropic assembly bias of the real-space groups. Note the negative sign of the quadrupole. The shaded regions in both panels represent the $1\sigma$ uncertainties computed as the error on the mean over 1000 simulations.
}
\label{fig:Rotation}
\end{figure*}

As noted above, the nonzero real-space quadrupole for redshift-space selected groups
arises in part from redshift-space selection effects (e.g., \citep{Chuang}) and in part from anisotropic assembly bias of real-space halos.  To help quantify which of these effects is dominant, we have attempted to remove any correlations between small-scale structure and large-scale tides before measuring the power spectrum. We do so by first finding all groups with $n_{\rm min} = 3$ and $b_{ll}=0.28$ in real-space, and then we randomly rotate each group.  
We subsequently put all particles in redshift-space, and find redshift-space groups as before. We then measure the quadrupole of the power-spectrum of these groups in both real- and redshift-space, and compare to the non-rotated versions in Fig.\ \ref{fig:Rotation}. In redshift-space (left panel) we see a large difference between the quadrupoles with and without rotation. Both have the same selection functions - being selected in redshift-space, but for one we have removed the physical correlation between small-scales and large-scales. This shows that a significant component of the quadrupole comes from halo assembly bias. Moving these groups to real-space (right panel) we see that, on scales much larger than the sizes of any of the rotated groups, the real-space quadrupole of the halo-matter cross-spectrum is significantly reduced: we have removed the effects of halo assembly bias, so only the redshift-space selection effects remain. The real- and redshift-space results are consistent, with an offset of similar magnitude from the Kaiser model prediction remaining in both, even after rotation. Such an offset is not surprising, even in the absence of halo assembly bias, due to the redshift-space selection effects \cite{Chuang}.

\section{Conclusions and summary}
\label{sec:Conclusions}

We have investigated the clustering of haloes found in redshift-space in order to help to understand the clustering of groups found in galaxy surveys, and the component of the clustering of galaxies and 21cm observations correlated with halo properties.

Using simulations, we have shown that assembly bias, the correlation (at fixed mass) between internal halo properties and large-scale environment, exists not only for scalar properties like age, concentration, etc., but also for non-scalar properties as well, like shapes, velocity dispersion, and angular momentum.  These correlations are not surprising, and have been known for some time: for example, tidal torque theory has long been hypothesized as the origin of angular momentum in galaxies and halos (e.g., \citep{Mo1998,Lee2001}), and the theory of ellipsoidal collapse \cite{BondMyers1996} predicts that halo shapes and dispersion tensors must be aligned with tidal fields in the sense observed in our simulations.  

Anisotropic assembly bias can impact on observed galaxy clustering in subtle ways. One example of this is already well-known in the cosmic shear literature: the correlation between tides and shapes leads to the GI alignment contaminant for weak lensing \cite{Hirata2004,Hirata2007}. As first pointed out by \citet{Hirata_th}, the radial projection of the GI correlation can similarly act as a contaminant for redshift-space distortions, whenever galaxy selection couples to galaxy shape. We have argued that similar contamination can arise whenever galaxy selection couples to LOS halo velocity dispersion or angular momentum. Provided that galaxy velocity dispersion and angular momentum correlate with those of their haloes, we could see an effect for spiral disk galaxies, when dust extinction and reddening are more severe for galaxies viewed edge-on than for face-on.  

Our work has more direct relevance for the analysis of groups of galaxies found in redshift surveys \cite{Eke04,Yang07}. For groups found in redshift-space, it is clear that the anisotropic clustering will be distorted from the Kaiser model by both assembly bias and redshift-space selection effects (for example a biased selection towards pairs of objects with a strong pairwise infall). Our results from simulations suggest that these effects could be of the same order as the standard RSD signal. Thus it may be difficult to use such group catalogues to split samples and use multi-tracer techniques to enhance the RSD signal \cite{mcdonald08}, or to weight galaxies to enhance the measured clustering \cite{percival04}, without also distorting that clustering away from the predictions of simple models.

At low redshift the 21cm signal observed by intensity mapping surveys like CHIME \cite{CHIME} is expected to originate primarily from dense gas clouds with column densities $N_{\rm HI} \gtrsim 10^{20} {\rm cm}^{-2}$ similar to damped Lyman-$\alpha$ absorbers \cite{Wolfe2005}.  At column densities $N_{\rm HI} \gtrsim 10^{21} {\rm cm}^{-2}$, 21cm can self-absorb, i.e.\ the sources can become optically thick \citep{2012ApJ...749...87B}. This can preferentially scatter 21cm photons away from observers in the galactic disk plane towards observers perpendicular to the disk plane, causing the 21cm signal seen by any observer to preferentially arise from face-on galaxies rather than edge-on galaxies.  As noted above, through the galaxy--halo correlation, this can generate a quadrupole that is indistinguishable from RSD, even though no actual galaxies are individually identified in the 21cm intensity map.  The presence of such contamination can severely complicate the interpretation of RSD observations. 
Conservatively, one might imagine that we will need to marginalize over an unknown linear tidal bias $b_q$ (see Eqn.\ \eqref{eqn:notKaiser_RSD}) much as we marginalize over an unknown linear bias $b$ when modeling observed galaxy correlations.  Doing so would degrade (or possibly lose) the information provided by measurements of the power spectrum quadrupole.  This degradation can be mitigated when we can place constraints on $b_q$ using independent measurements.  For example, as explained by \citet{Hirata_th} in the context of galaxies, the same correlation between tidal fields and radial alignment that contaminates galaxy RSD also appears as a correlation with transverse alignments, meaning that measurements of the GI correlation for any particular galaxy sample can help to limit the magnitude of $b_q$ for that sample \cite{Hirata_obs}.  This does require us to resolve individual sources, and so this mitigation procedure may not be possible for intensity mapping surveys with low angular resolution.

We can also view this effect more positively as an opportunity to detect halo assembly bias.  Scalar (isotropic) halo assembly bias has eluded observational detection \cite{Sunayama2019} despite considerable effort \cite{Lin2016,Miyatake_assembly_evidence} as noted above, in part since it is degenerate with halo mass uncertainty.  In contrast, anistropic assembly bias cannot be mimicked by varying the halo mass.  Using existing galaxy redshift surveys, it may be possible to detect various forms of this assembly bias.  For example, we could imagine comparing the redshift-space clustering of galaxies of similar stellar mass but different 1D (LOS) velocity dispersion.  Similarly, our results suggest that the RSD of groups selected in redshift-space will have enormous deviations from standard Kaiser RSD, analogous to predictions for void clustering \cite{Chuang}.  Such effects may be detectable already using existing data from BOSS, or from upcoming DESI data.

\begin{acknowledgments}
We thank Francisco Villaescusa-Navarro for providing us with full outputs of the Quijote simulations, and Phil Mansfield for sending us halo catalogs from the $\nu^2$GC-L simulation. We also thank Niayesh Afshordi, Emanuele Castorina,  Albert Chuang, Joanne Cohn, Andreu Font-Ribera, Gil Holder, Faizan Mohammad, Hamsa Padmanabhan, Uro\v{s} Seljak, Marko Simonovi\'{c}, Kendrick Smith, and Martin White for useful dicussions. This research was supported by the Centre for the Universe at Perimeter Institute.  
Research at Perimeter Institute is supported in part by the Government of Canada through the Department of Innovation, Science and Economic Development Canada and by the Province of Ontario through the Ministry of Economic Development, Job Creation and Trade. We also acknowledge support provided by Compute Ontario (www.computeontario.ca) and Compute Canada (www.computecanada.ca). We acknowledge the use of \texttt{nbodykit} \cite{nbodykit}, \texttt{Pylians}\footnote{\href{https://github.com/franciscovillaescusa/Pylians}{https://github.com/franciscovillaescusa/Pylians}}, \texttt{IPython} \cite{IPython}, \texttt{Matplotlib} \cite{Matplotlib} and \texttt{NumPy}/\texttt{SciPy} \cite{Numpy}.
\end{acknowledgments}
\bibliography{References.bib}

\end{document}